\documentclass{emulateapj}
\shorttitle{Wide-separation planets in Upper Scorpius}
\shortauthors{Ireland et al.}
\begin{document}

\title{Two Wide Planetary-Mass Companions to Solar-Type Stars in Upper Scorpius.}

\author{Ireland, M.J.}
\affil{Sydney Institute for Astronomy (SIfA), 
  School of Physics, University of Sydney NSW
  2006, Australia}
\author{Kraus, A. \footnote{Hubble Fellow}}
\affil{Institute for Astronomy, University of Hawaii, 2680 Woodlawn
  Drive, Honolulu, HI 96822, USA}
\author{Martinache, F.}
\affil{National Astronomical Observatory of Japan,
  Subaru Telescope, Hilo, HI 96720, USA}
\author{Law, N.}
\affil{Dunlap Institute for Astronomy and Astrophysics, University of
  Toronto, 50 St. George Street, Toronto M5S 3H4, Ontario, Canada}
\and
\author{Hillenbrand, L.A.}
\affil{California Institute of Technology, Department of Astrophysics,
  MC 105-24, Pasadena, CA 91125}

\newcommand{\pfifty}{1RXS J160929.1-210524}
\newcommand{\pseventy}{GSC 06214-00210}
\newcommand{\kninenine}{1RXS J160703.4-203634}

\label{firstpage}

\begin{abstract}
At wide separations, planetary-mass and brown dwarf companions to
solar type stars occupy a curious region of parameters space not
obviously linked to binary star formation or solar-system scale
planet formation. These companions provide insight into the extreme
case of companion formation (either binary or planetary), and due to
their relative ease of observation when compared to close companions,
they offer a useful template for our expectations of more typical
planets. 
We present the results from an adaptive optics imaging survey for wide
 ($\sim$50-500\,AU) companions to solar type stars in Upper Scorpius. We
 report one new discovery of a $\sim$14\,M$_{\rm J}$ companion around
 \pseventy, and confirm that the candidate planetary mass companion
 \pfifty~detected by \citet{Lafreniere08} is in fact co-moving with its primary
 star. In our survey, these two detections correspond to $\sim$4\% of solar type 
 stars having companions in the 6-20\,$M_J$ mass and $\sim$200-500\,AU
 separation range. This figure is higher than would be
 expected if brown dwarfs and planetary mass companions were drawn
 from an extrapolation of the binary mass function. Finally, we  discuss
 implications for the formation of these objects.
\end{abstract}
\keywords{brown dwarfs, planetary systems, Infrared: planetary systems, }

\section{Introduction} 
\label{secIntro}
Over the past five years, direct imaging surveys for extrasolar
planets have discovered a small but significant number of ultra-low
mass companions (henceforth ULMCs, masses $\la$20\,$M_{\rm J}$) at $\ga$50 AU separations from their
primaries. 
These objects have estimated masses that are in the same range as 
radial velocity or transiting planets with $<$5\,AU separations, which have a
continuous mass distribution up to $\sim$20\,$M_{\rm J}$, then a gap
in mass until arguably star-like objects are found at $>$60\,M$_J$
\citep{Grether06,Deleuil08,Bouchy10,Anderson10}. For this reason,
ULMCs are often called ``planetary mass'' companions.

The prototypical wide ULMC, 2M1207-3933, consists of a 4--8
$M_{Jup}$ companion located $\sim$50 AU away from a 10 Myr old brown
dwarf \citep{Chauvin04}. Since its discovery, half a dozen other ULMCs
have also been reported, most of which orbit much higher-mass
primaries ($\sim$0.5--2.0 $M_{\odot}$;e.g.   
\citealt{Lafreniere08}; \citealt{Kalas08}; \citealt{Marois08}). Some of these
systems appear to be genuinely scaled-up versions of our own solar
system, with planets that are consistent with formation in a disk. 
One example is the companion to Fomalhaut, which is coplanar with its
debris disk. Another is the HR~8799, which has 
multiple planetary companions of
similar mass. Other cases like CHXR~73 and
\pfifty~are more ambiguous since their orbital radii are even
wider, and it is unclear whether they lie in the original plane of
planet formation. Such companions could very well form like planets
within a circumstellar disk or like binaries from the collapse of a
molecular cloud.

These ULMCs pose a significant challenge to existing models of planet
and binary formation. Their orbital radii are so large that the core
accretion timescale ($>$100 Myr at 100 AU; Pollack et al. 1996) should
be much longer than the typical protoplanetary disk dissipation
timescale ($\sim$3-5 Myr; \citealt{Haisch01}; 
\citealt{Hernandez07}; \citealt{Currie09}).
Some of the closer companions could potentially form on the
gravitational instability timescale \citep[e.g.][]{Boss01}, which can be
very short at intermediate radii (10--100\,AU), but gravitational instability in disks has
not been extensively modeled at $\ga$100 AU, much less at $\sim$300 AU. These
companions could represent the extreme end of the binary mass
function, which appears to be linearly-flat (with all companion masses
being equally probable, \citealt{Kraus08,Raghavan10}) well into the substellar regime. However, it
is unclear whether this trend could extend to planetary masses
(by which we mean $<$20\,M$_{\rm J}$ in this paper). These
companions have masses near the opacity-limited minimum mass
\citep{Hoyle53,Low76,Bate05}, and unless their formation occurred exactly as the
circumstellar envelope was exhausted, then they should have quickly accreted
enough mass to become high-mass brown dwarfs or stars.

In this paper, we report the discovery of two ULMCs in the Upper
Scorpius OB association, one of which was independently discovered by
\citet{Lafreniere08}. The survey consists of a subset of the
aperture-masking interferometry sample reported by
\citet{Kraus08}.
We describe our observations and data analysis techniques in Section
\ref{sectObs}, and in Section~\ref{sectFollowup}, we report the
detections and detection limits from our survey.
In Section~\ref{sectFreq}, we report the first measurement of a
frequency for ULMCs around young stars. Finally, in
Section~\ref{sectDiscuss}, we discuss the implications for possible
formation mechanisms for ULMCs.

\section{Observations}
\label{sectObs}

\subsection{Discovery Observations}
\label{sectDisco}

Nearby ($\la$200\,pc) young ($\la$20\,Myr) stars have been the object of numerous high-resolution
imaging campaigns over the past several decades. These observations
have included lunar occultation (e.g. \citealt{Simon95}), speckle
interferometry (e.g. \citealt{Ghez93}), Adaptive Optics (AO) imaging
(e.g. \citealt{Lafreniere07,Masciadri05,Chauvin10}), and most recently
non-redundant masking (NRM) inteferometry \citep{Kraus08}. 
The earlier techniques of lunar occultaion and speckle were only sensitive to the presence of
bright, stellar-mass, companions, and only recently did AO imaging and
NRM interferometry managed to probe within the brown dwarf regime,
going as far as sampling the top of the planetary mass regime.

In K08, we reported the results of one such survey of
young stars in the Upper Sco OB association that used a combination
of conventional AO imaging and NRM-interferometry.
In addition to the results for stellar and brown dwarf companions,
that paper also reported detection limits for ULMCs at
small separations (within 50 AU) where the probability of background
star contamination was negligible.
In this work, we report the corresponding analysis for candidate
companions at wide separations, including multi-epoch follow-up imaging
for three candidates, of which two appear to be associated and of
approximately planetary mass.

Table 4 of K08 lists the AO imaging observations conducted with the 
PHARO camera at the Palomar 200" telescope and the NIRC2 camera at the
Keck-II 10m telescope. We found that 10 out of these 62 targets had
stellar binary companions at separations of $\sim$0.25--5.0\arcsec, which
should mean that the majority of additional faint companions are not
dynamically stable in this range of separation; these were omitted from our sample.
In addition, we omitted the few targets with spectral types of later than
M2 ($M<0.5 M_{\sun}$) so as to work with a single mass range of
approximately ``solar-type'' stars, leaving 49 targets in 
consideration.

All observations used the smallest pixel scales (10 mas/pix with NIRC2
and 25 mas/pix with PHARO) in order to achieve the best PSF
sampling. We used a $K_s$ filter at Palomar and the  $Br\gamma$ filter
at Keck, which yielded diffraction-limited resolutions of 100\,mas and
50\,mas, respectively. The Keck observations used the narrowband
filter despite the penalty in sensitivity, because the brighter primary
stars in our sample
would have saturated the detector within the minimum exposure
time. Much of this sensitivity was regained by using
 more Fowler samples per frame, which significantly reduces the
high read noise of NIRC2's detector.
These observations used relatively short total integration time, of the
order of a minute, in comparison with other AO imaging surveys
(e.g. \citealt{Metchev09}), resulting in limiting magnitudes of 
$K_{lim}\sim$15--17 for the 
companions. Given the young age of these Upper Sco targets ($\sim$5 Myr), even
these shallow observations were able to reach the planetary mass
regime (at 2 arcsec separations, $\la$14  $M_{Jup}$ for 48 targets and $\la$7 $M_{Jup}$
for 42 targets).

The detections and detection limits were derived using methods we
previously described in \citet{Krausthesis09} and Kraus \& Hillenbrand
(2010, in prep). Since the detection limits at more than a few $\lambda /D$ are
driven by speckle noise, detections and detection limits were
determined from the coadded image stacks by placing a large number of
photometric apertures (with diameter $\lambda$$/D$) around each
target, then measuring the mean and standard deviation of the
brightness distribution for all apertures in concentric rings around
the primary. For each ring, we identified all candidate detections
with significance $\ga$5$\sigma$ above the mean brightness, then
compared those detections to other stars taken on the same night to
identify and reject the quasistatic speckles and diffraction spikes
that are seen in common for many targets. All candidate detections
that could not be identified as PSF artifacts were then adopted as
genuine sources and hence as candidate bound companions, while the
5$\sigma$ limits were adopted as our formal detection limits. At large
separations, the detection limit from this algorithm converges to the
sky background limit, and at these separation ($>$2\, arcsec), we used
a 10$\sigma$ limit, because of the large number of pixels in this
regime. We found no candidate companions near the
detection limits in the speckle-limited regime, but there are many
candidate companions in the sky-limited regime. 

We determined the photometry and astrometry for these sources using
the methods described in K08 and \citet{Kraus09}. To
briefly summarize, we measured astrometry and aperture photometry for
each source with respect to the known USco member using the IRAF task
DAOPHOT \citep{Stetson87}; all measurements were conducted using
apertures of 0.5, 1.0, and 2.0 $\lambda /D$, and then the optimal
aperture was chosen to maximize the significance of the detection
(given the competing uncertainties from the sky background and the
Poisson noise for the source itself). In order to estimate the
uncertainties from the data, we analyzed the measurements in
individual frames and then combined those measurements to estimate the
mean and standard deviation. We then accounted for the systematic
uncertainty in the plate scales and distortion solutions of PHARO
(0.3\%; Section 2.3) and NIRC2 (0.05\%; Ghez et al. 2008,
\citealt{Cameron08}) by adding
those terms in quadrature with the observed scatter. Most of the new
candidate companions were identified with PHARO; as we describe below,
its astrometric calibration is not yet well-understood and might be
further improved with more calibration observations, but our
systematic uncertainties should account for this effect. Many of the
wider ($\ga$5\arcsec) companions are also affected by anisoplanatism,
yielding systematically low brightness estimates for aperture
photometry. A correction of the photometry would require a
detailed knowledge of the atmospheric turbulence profile and
isoplanatic patch size, so it can not be accomplished for our
data. However, as we describe below, all of these sources have
well-calibrated $K$ magnitudes available from UKIDSS. We choose
instead to defer to those measurements in determining colors (Section
3.2). 

Finally, all candidate companions with separations $\ga$4.2\arcsec\,
have spatially resolved counterparts in the UKIDSS Galactic Cluster
Survey \citep{Lawrence07}, and many of the widest companions also
have optical counterparts in the USNO-B1.0 digitization of the Palomar
Observatory Sky Survey \citep{Monet03}. As we describe further in
Section 3.2, we have used those observations to identify most of these
candidate companions as background stars. The rest require multi-epoch
astrometric monitoring to determine whether they are associated. 

\subsection{Follow-up Observations}
\label{sectFollowup}

\begin{figure}
\plotone{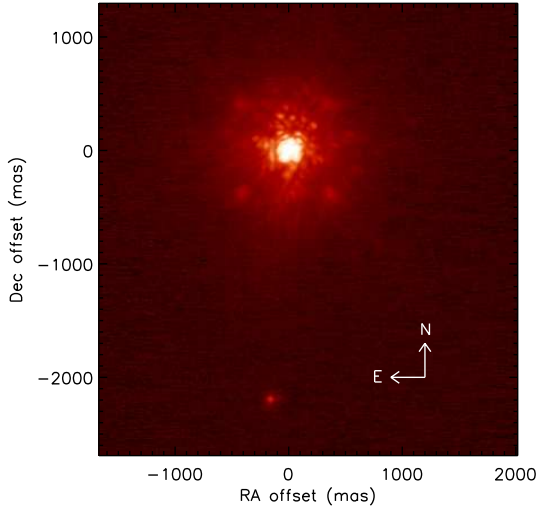}
\caption{An example image of \pseventy~from 2008 July in the Kp
  filter, with a log stretch. The faint companion can be seen as at an
RA offset of $\sim$-200\,mas and a Dec offset of $\sim$-2200\,mas.}
\label{figImage}
\end{figure}

Around three targets in the K08 survey we found faint visual candidate
companions with projected separations of 2--3\arcsec\, and
brightnesses of $K\sim$15--16: \pseventy, \pfifty, and \kninenine,
whose basic properties are listed in Table~\ref{tabprops}. We re-observed
these objects with the telescope-instrument-filter combinations as
detailed in Table 2. 
 
The expected surface density of unassociated background stars at
these brightnesses is relatively low, but since Upper Scorpius is at moderate
galactic latitude and is projected 
over the background Milky Way bulge, this probability was not
sufficiently low that we could assume they were bound ULMC
companions. Follow-up observations were therefore required to confirm
common proper-motion.

The
density of background stars at $K<17$ is approximately 0.002 per
square arcsecond (Kraus \& Hillenbrand 2010, in prep.), which
gives a 86\% chance of a chance alignment within 2.5\arcsec for at
least one star in our sample, but only a 30\% chance of all 3 candidates
being chance alignments. 
It is impossible to use proper motions to rule out chance alignments
with other association members because the internal velocity
dispersion of Upper Sco is only $\sim$1\,km\,s$^{-1}$
($\sim$1.5\,mas\,yr$^{-1}$, \citealt{Kraus08b}),
meaning that they would be comoving even if they were only seen in
chance alignment. However, 
the surface density of young stars in Upper Sco
is no more than 100\,deg$^{-2}$ \citet{Kraus08b} or 
$10^{-5}$\,arcsec$^{-2}$, meaning that the probability of a chance alignment is
negligible. 

The proper motion of the Upper Scorpius association is (-11.5,-23.5)\,mas\,yr$^{-1}$ in
equatorial coordinates, determined from the members listed in
\citet{deZeeuw99}, and using updated proper motions from
\citet{vanLeeuwen07}.  This is relatively small when
compared to nearby moving groups, so an
astrometric accuracy of order 2\,mas or 0.1\% at 2\arcsec is required
in order to clearly determine whether a companion is co-moving or not
on a 1 year time baseline.

The NIRC2 camera has been shown to have a stability better than this
in the precise galactic center astrometric work of \citet{Ghez08}.
\cite{Cameron09} showed that the PALMAO adaptive optics system
\citep{Troy2000} and the PHARO Near Infrared (NIR) camera are stable in distortion to
$\sim$100$\mu$as over several months. 
However, PALMAO underwent several upgrades between these common proper
motion confirmation observations, including preliminary work for a
reconfiguration of the output beam path to accommodate new science
instruments.

We searched for PALMAO distortion solution changes during this period
using observations of the core of the M5 globular cluster from four
nights (28/5/2007; 29/5/2007; 17/07/2008 and 17/08/2008). The
observations consisted of 100 co-added 1.4 second exposures at each
epoch, were taken in the 25 mas PHARO plate scale, used the
Br-$\gamma$ narrow-band filter to avoid differential chromatic
refraction, and were timed to be observed at very similar airmasses
and hour angles. The same AO guide star was used in each case, and
care was taken to align each epoch to the same pointing within an
arcsecond and to keep all other AO and camera parameters consistent.
We fitted 2D Gaussians to derive the positions of 34 stars covering
the 25"$\times$25" field in each dataset; we also checked the fitted
positions using SExtractor \citep{Bertin1996} and found very similar
results.

We first used a simple model to match the stellar positions between
epochs, allowing for a change in pointing position, an arbitrary
rotation, and a separate scaling in the X and Y axes. There was no
detectable change in rotation or scale within 2007 or within 2008, but
between those years the field rotation changed by 0.18 degrees, while
the X and Y plate scales changed by -0.5 \% and +0.2\%
respectively. However, the simple rotation and plate scale change model
leaves 3-5mas residuals in matching the 2007 and 2008 measured
positions. The residuals are reduced to $\sim$1mas using a 
general 4-parameter linear transformation, where the additional
parameter represents the plate scale 
changes being in arbitrary orthogonal axes, or alternatively the
addition of a shear term. However, some statistically significant
higher-order distortion is still apparent after application of this
solution. It appears that long-term precision astrometry using
adaptive optics requires careful attention to changes in the detailed
distortion solutions (beyond simple rotation and scale); this is
particularly important as, like PALMAO, the optics of many adaptive
optics systems are regularly upgraded.

Therefore, wherever at least two NIRC2 epochs are available, we
only consider the astrometry with NIRC2 for our proper motion
measurements. All followup NIRC2 data were acquired using the
narrow camera with a $\sim$10\,mas/pixel image scale and the $K_p$
filter, which we found did not quite saturate for these targets using
the shortest exposure time with a minimum number of Fowler samples.

The follow-up NIRC2 data were reduced using a custom pipeline written in
IDL. After standard image-processing tasks (background subtraction,
flat-fielding and bad pixel removal) were completed, the
distortion in the image plane was then removed by the program {\tt
nirc2dewarp} available from the NIRC2 camera home page, using the
updated distortion solution of B.Cameron 
(personal communication, 2007). Finally, aperture-photometry and
centroids were computed with the {\tt ImExam} function of {\tt atv},
version 2.0b4. Selected data sets were also analysed with the DAOPHOT
function of IRAF, giving results consistent well within the 
error bars. 

For centroiding, a centering box size of 5 pixels was used, with the
exception of one single observation in L-band, where a box size of 9
pixels was used.
The  aperture-radius for photometry was 5 pixels for the Kp filter,
but 7 pixels for the J and H filters due to low Strehls and dispersion, and
10 pixels in L-band due to the larger diffraction-limited core.
The sky annulus was always set at 10 to 20 pixels. Note that we did
not attempt to calibrate absolute photometry, and have only computed
the relative photometry between primary and secondary.

\section{Results}
\label{sectInd}

The following subsections describe the multi-epoch astrometry obtained
on each of the three targets in Table~\ref{tabprops}, in order to confirm proper motion. Two
clearly associated objects are discussed, as well as two objects
identified as background stars.

\subsection{New Companions}

\begin{deluxetable*}{lccccrr}
\tabletypesize{\scriptsize}
\tablewidth{0pt}
\tablecaption{Stellar Properties}
\tablehead{
\colhead{Name} & \colhead{RA} & \colhead{DEC} & 
\colhead{SpT} & \colhead{Mass} & 
\colhead{$R$} & \colhead{$K$}
\\
\colhead{} & \multicolumn{2}{c}{(J2000)} & \colhead{} & 
\colhead{($M_{\sun}$)} & \colhead{(mag)} & \colhead{(mag)}
}
\startdata
GSC 06214-00210      &16 21 54.67&-20 43 09.1&M1&0.60&11.6&9.15\\
1RXS J160929.1-210524&16 09 30.30&-21 04 58.9&K7/M0&0.68-0.77&12.1&8.92\\
1RXS J160703.4-203634&16 07 03.56&-20 36 26.5&M0+?&0.68+0.59&11.3&8.10\\
\enddata
\tablecomments{$R$ magnitudes are from USNO-B, while coordinates and
  $K$ magnitudes are from 2MASS. The mass of the secondary star for
  1RXS J160703.4-203634 is inferred from the mass ratio, while other
  masses are directly inferred from the mass-temperature relations of
  Baraffe et al. (1998). Spectral types are taken from the discovery
  sources, Preibisch et al. (1998) and Kunkel (1999).} 
\label{tabprops}
\end{deluxetable*}

\subsubsection{\pseventy\,b}

\pseventy\, is a young M1 star near the eastern edge of Upper
Sco. It was originally identified as a candidate young star by
Preibisch et al. (1998) based on its X-ray emission, then confirmed to
have strong lithium absorption ($EW=0.38 \AA$) and H$\alpha$ emission
($EW=-1.51 \AA$) as compared to stars of equivalent spectral type in
several young clusters. It was also found to have a proper motion
$\mu=(-18.6\pm 1.7,-32.2\pm 1.7)$ mas yr$^{-1}$ \citep{Zacharias10}, 
consistent with the motion of Upper Sco.
Given its M1 spectral type, the 5 Myr mass-temperature relations
of \citet{Baraffe98} predict a mass of 0.60\,$M_{\sun}$. Note that
this mass must be taken with caution, as an uncertainty of 1 subclass
in spectral type should correspond to an uncertainty of $\sim$0.1
$M_{\sun}$ in mass. One should keep in mind that the models themselves
carry an unknown uncertainty since the mass-luminosity and
mass-temperature relations of young stars are almost completely
uncalibrated for $\la$1 $M_{\sun}$ (e.g. \citealt{Hillenbrand04}). 

\begin{figure}
\epsscale{0.8}
\plotone{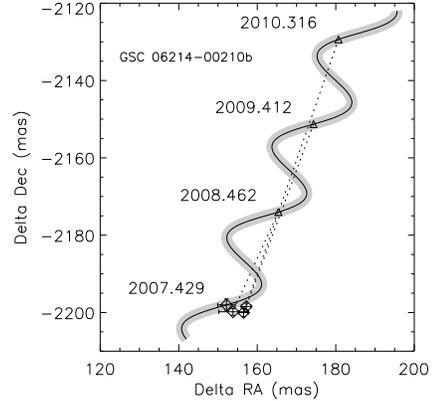}
\caption{The observed position of the companion to \pseventy (diamonds), with the
expected motion of the companion if it were a background star with
respect to the first epoch
over-plotted (solid line, triangles at the times of observation).  The
dashed lines join the locations of the companion at each epoch with the expected
position if it were a background star. This
shows that \pseventy b is physically associated. }
\label{figP70}
\end{figure}

An example image of \pseventy b used for astrometry is shown in
Figure~\ref{figImage}. Our photometric and astrometric observations of
this companion are
given in Table \ref{tabP98-70} and Figure \ref{figP70}.
The candidate companion showed a relative motion of 7$\pm$4
mas with respect to \pseventy\, over 2.9 years. This is much less
than the $\sim$70 mas motion expected if \pseventy b were a background
star, and demonstrates that it is most likely physically associated
with \pseventy.
The apparent relative proper motion of 2.5$\pm$1.3\,mas\,yr$^{-1}$
corresponds to 1.7$\pm$0.9 km\,s$^{-1}$ assuming a 7 mas parallax for Upper
Scorpius, which is roughly equal to the circular orbital velocity of
1.7\,km\,s$^{-1}$ expected for a $\sim$300 AU orbit. 

As can be seen from the larger contrast in bluer filters (cf. Table
\ref{tabP98-70}), \pseventy b is quite red compared to its primary
star. The primary has an observed 2MASS color of $J-K=0.85\pm0.04$,
and most M1 stars have a typical $K-L'$ color of $\sim$0.15$\pm$0.05
\citep{Leggett92}, so the inferred colors for the companion are
$J-K=1.35 \pm 0.11$ and $K-L=1.18 \pm 0.10$.
The mean distance for Upper Sco is 145 pc (de Zeeuw et al. 1999), and
a $\pm 6$\, degree spread on the sky of the densest region (where
  all our candidates belong) corresponds to a $\sim$14\,pc error on
  the distance of any individual star. This gives a
a distance modulus of $m-M=5.8 \pm 0.2$, so given the
2MASS magnitudes of the primary, the absolute magnitudes of the
companion are $M_J\sim 10.5$, $M_H\sim  9.6$, $M_K\sim 9.1$, and $M_{L'}\sim
7.9$. The interstellar extinction toward Upper Sco is negligible
($A_V\la 1$ or $A_K\la 0.1$), so no extinction corrections should be
required.

The observed $J-K$ color of 1.3 is consistent with the M8-L4 spectral
type range of field dwarfs from \citet{Leggett02}. The K-L’ color
of 1.05, however, would clearly place the star at the red end of this
range, at L3-L4. The absolute K-magnitude of 9.1 is then about 2
magnitudes brighter than for corresponding field objects, providing
clear evidence of a larger radius and young age; this height above the
main sequence is similar to that observed for other late-type members
of Upper Sco \citep{Lodieu07} and TW Hya \citep{Mamajek05,Teixeira08}. 

\begin{figure}
\epsscale{0.8}
\plotone{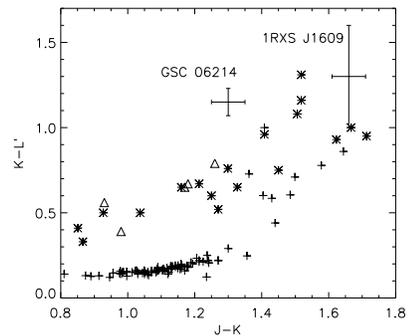}
\caption{J-K versus K-L' color for our two confirmed companions, with
  the colors overplotted for M dwarfs from \citet{Leggett92}
  (triangles),M4 to L6 dwarfs from \citet{Golimowski04} (asterisks),
   and for M giants from \citet{Fluks94} (crosses). 
The L band photometry of \pfifty comes from \citet{Lafreniere10}.}
\label{figColorcolor}
\end{figure}

The colors and spectral type of \pseventy b are
not expected to match field dwarfs due to the low surface
gravity. \citet{Allers10} find that J-K and K-L’ colors are
significantly redder for low gravity objects at a given spectral
type. Although \pseventy b is a low-mass object, \citet{Allers10}
focusses on redder object at J-K colors of $\sim$2.0, and it isn't
clear that their result should hold for bluer
objects like \pseventy b, with J-K=1.3. In Figure~\ref{figColorcolor}
we show the NIR
the colors of \pseventy b with respect to field dwarfs and giants
(which bracket \pseventy b in gravity),
where it appears to have a small but significant K-L' excess. The red
K-L’ color could be due to a disk, as $\ga$1/3 of young brown dwarfs
retain a disk for $\ga$5 Myr (e.g. \citealt{Scholz07}), with a trend
of increasing disk lifetime with decreasing mass. Spatially
resolved spectroscopy of the system would be required in order to
determine the spectral type more accurately, and to search for signs
of accretion onto the secondary.

Given the observed brightness in the $JHK$ filters, a comparison to
the 5 Myr DUSTY models \citep{Chabrier00} et al. 2000) suggests that 
the mass of the companion is $\sim$12--15
$M_{Jup}$, with bluer filters suggesting slightly lower masses than redder
filters.
The COND models \citep{Baraffe03} 
never predict sufficiently red $J-K$ colors to match our observations,
and indeed are not appropriate for objects with $T>1300$\,K.
This trend qualitatively matches the observed trend for young
low-mass objects to be redder than older field counterparts of similar
spectral type. If the comparison was based only on observed J-K color,
then the DUSTY models would predict masses of 10--12 $M_{Jup}$. 
As we discussed above, the $L'$ photometry could
have an excess from a circumstellar disk, so we suggest that the $L'$
magnitude should not be used directly in the mass estimate.

\begin{deluxetable}{llrrr}
\tabletypesize{\scriptsize}
\tablewidth{0pt}
\tablecaption{Followup Observations}
\tablehead{
\colhead{Julan Date} & \colhead{Band} &
\colhead{Separation} & \colhead{Position Angle} &
\colhead{Contrast} \\
\colhead{} & \colhead{} &
\colhead{(mas)} & \colhead{(degrees)} & \colhead{(mags)}
}
\startdata
\multicolumn{3}{l}{\em \pseventy\,b}\\
 2454258.0 & Kp & 2203.3$\pm$1.5 & 176.04$\pm$0.06& 5.74$\pm$0.05\\
 2454634.8 & Kp & 2204.7$\pm$0.9 & 175.99$\pm$0.03& 5.78$\pm$0.03\\
 2454634.8 & J  & 2205.2$\pm$0.9 & 176.00$\pm$0.09& 6.30$\pm$0.03\\
 2454982.9 & Kp & 2204.1$\pm$0.9 & 175.91$\pm$0.03& 5.88$\pm$0.03\\
 2455313.1 & Kp & 2205.6$\pm$1.1 & 175.93$\pm$0.03& 5.73$\pm$0.03\\
 2455313.1 & H & 2202.8$\pm$2.2 & 175.91$\pm$0.04& 6.21$\pm$0.03\\
 2455313.1 & L' & -\tablenotemark{a} & -\tablenotemark{a} & 4.75$\pm$0.05\\
\multicolumn{3}{l}{\em \pfifty\,b}\\
 2454634.8 & Kp & 2210.1$\pm$1.0 &  27.62$\pm$0.04 & 7.27$\pm$0.02 \\ 
 2454982.9 & Kp & 2211.3$\pm$0.9 &  27.61$\pm$0.05 & 7.23$\pm$0.03 \\ 
\multicolumn{3}{l}{\em \pfifty\,c}\\
 2454250.8 & Ks & 4261$\pm$14 & 219.5$\pm$0.2   & 8.63$\pm$0.04 \\ 
 2454982.9 & Kp & 4215$\pm$5  & 220.03$\pm$0.08 & 8.6$\pm$0.1 \\
\multicolumn{3}{l}{\em \kninenine\,b}\\
 2454634.8 & Kp & 2143.2$\pm$1.5 & 234.01$\pm$0.07 & 8.20$\pm$0.08 \\
 2454634.8 & J & 2138.2$\pm$1.0 & 234.05$\pm$0.07 & 8.06$\pm$0.05\\ 
 2454982.9 & Kp & 2123.1$\pm$0.5 & 234.45$\pm$0.01 & -\tablenotemark{b} \\ 

\enddata
\tablenotetext{a}{Data not suitable for precision astrometry.}
\tablenotetext{b}{Data were taken through the {\tt corona400} coronograph at
1.04$\pm$0.03 mags contrast and $\sim$7\,mags extinction of the
primary (not
precisely calibrated).}
\label{tabP98-70}
\end{deluxetable}

\subsubsection{\pfifty\,b}
\label{sectPFifty}

\pfifty\, is a young M0 star located near the center of Upper Sco. 
Like \pseventy, it was first identified as a likely Upper Sco member
by Preibisch et al. (1998) and exhibits both strong lithium absorption
($EW=0.54 \AA$) and H$\alpha$ emission ($EW=-1.14 \AA$). The proper
motion reported by UCAC3 \citep{Zacharias10} for this object is
(-11.2,-21.9) $\pm$1.5 mas yr$^{-1}$, which is also consistent with
the value for Upper Sco.
The spectral  type reported by Preibisch et al. (1998) was M0, but, as
was described by \citet{Lafreniere08}, newer measurements by
D. C. Nguyen suggest a more likely spectral type of K7. The inferred
mass would be 0.68-0.77 $M_{\sun}$ for the two estimates, with
uncertainties similar to that for \pseventy.

\begin{figure}
\epsscale{0.8}
\plotone{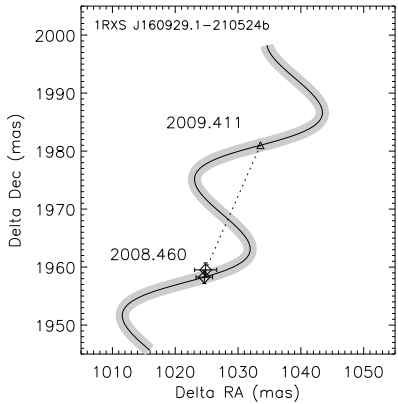}
\caption{The observed position of the closer companion to \pfifty, with
  symbols as in Figure~\ref{figP70}. The two observations are at nearly
indistinguishable locations in this plot, demonstrating that the
companion is physically associated..}
\label{figP50}
\end{figure}

Our observations for \pfifty b are listed in Table \ref{tabP98-70} and
plotted in Figure \ref{figP50}. The companion was first detected in
Palomar images as early as May 2007, but as the astrometric performance
of the PHARO camera was unverified on large timescales (cf. Section
\ref{sectFollowup}), we were not able to convincingly demonstrate
common proper motion until 2009. 
This object was independently detected by \citet{Lafreniere08}
and further characterized by \citet{Lafreniere10}. Our astrometric 
results are consistent with those recently reported by
\citet{Lafreniere10} at 2-$\sigma$ in separation and at 1-$\sigma$ in
position angle, but have higher precision due to the accurate
astrometric characterization of NIRC2.
Based on a spectrum indicating a late spectral type and low gravity,
these authors argued that this faint companion was a young object
and therefore likely a physical companion.
Since that time, we have continued monitoring this object (cf. Table
\ref{tabP98-70}, and can now confirm that \pfifty b is a physical
companion of \pfifty. The apparent proper motion of
1.2$\pm$1.3\,mas\,yr$^{-1}$, corresponding to 0.8$\pm$0.9\,km\,s$^{-1}$. This is
consistent with the $\sim$1.7\,km\,s$^{-1}$ orbital motion expected,
especially considering possible projection effects.

Since \citet{Lafreniere08} already reported NIR colors, we only
took $K$-band observations. These measurements are consistent with
theirs, and indicate an absolute magnitude of $M_K\sim 10.4$. 
The predicted mass from the DUSTY and COND models is $\sim$8 $M_{Jup}$
from the $K$ band photometry or $\sim$7 $M_{Jup}$ from $J$ band
photometry, though again, these estimates are completely uncalibrated
by observations. Given the assumptions listed above for the distance
of Upper Sco, the projected separation is $\sim$320 AU, similar to the
value for \pseventy\, system.

\begin{figure}
\epsscale{0.8}
\plotone{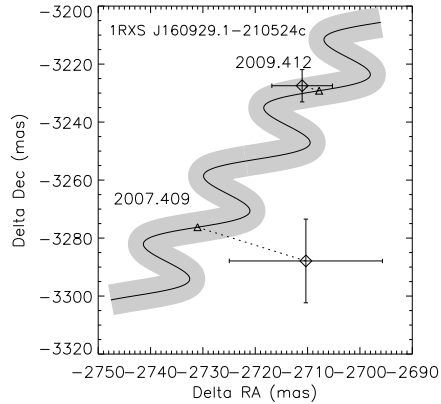}
\caption{The observed position of the wider companion to \pfifty, with
  symbols as in Figure~\ref{figP70}. As the first epoch has such large
  errors, we reference the apparent motion expected from a background star to
  the weighted average of the two epochs rather than to the first
  epoch. Motion of \pfifty with respect to this object is clearly
  detected, demonstrating that the candidate companion is a background
star.}
\label{figP50c}
\end{figure}

\subsection{Background Stars}

Unassociated background stars typically are identified based on
multi-epoch astrometric monitoring (which indicates that they are not
comoving) or multi-wavelength observations to measure colors (which
indicate that they do not fall along the same color-magnitude
sequence). As we describe in the next several subsections, several of
the closer companions will require astrometric monitoring with
high-resolution imaging to confirm or disprove their association, but
the wider companions can be identified and rejected based on archival
photometry from seeing-limited all-sky surveys. 

\subsubsection{\pfifty\,c}

In addition to the companion listed as companion b in Table~\ref{tabP98-70}, an
additional wider companion candidate, listed as companion c in
Table~\ref{tabP98-70} was found around \pfifty~in both the 2007 Palomar images
(see Section~\ref{sectWider}) and the 2009 Keck images. This companion
candidate was also reported by \citet{Lafreniere08}. Although the
Palomar astrometry had significant errors, the time baseline of 2
years was sufficient to clearly show that this object was a background
star, as shown in Figure~\ref{figP50c}.

\subsubsection{\kninenine\,b}

\kninenine\, is a young, close binary system which is also located
near the center of Upper Sco. It was originally identified in
a survey for new Sco-Cen members by \citet{Kunkel99}, using methods very
similar to the survey by Preibisch et al. (1998): targets were
identified as potential X-ray emitters, then confirmed to have strong
lithium for their age. However, the survey by Kunkel was never
published in the refereed literature, so the equivalent widths of relevant spectral
lines are not available. We discovered its multiplicity during our
direct imaging and aperture masking survey (K08), finding that it had
a companion with projected separation of 184 mas ($\sim$27 AU) and
flux ratio $\Delta K = 0.15$ (mass ratio $q\sim 0.87$). The same
observations also revealed a faint candidate companion at a projected
separation of $\sim$2.1\arcsec. 

\begin{figure}
\epsscale{0.8}
\plotone{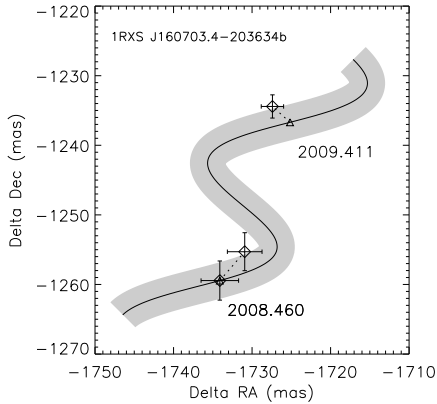}
\caption{The observed position of the companion to \kninenine~with
  respect to the center of light of the close binary, , with
  symbols as in Figure~\ref{figP70}. This candidate companion is consistent with
  a background star.}
\label{figK99}
\end{figure}

As we summarize in Table \ref{tabP98-70} and Figure~\ref{figK99},
multi-epoch astrometry for the candidate companion shows that it is
not comoving with \kninenine, but instead appears to be nearly
stationary, as would be expected for a background star. The relative
motion of the primary with respect to the companion is
(7$\pm$2,-26$\pm$2)\,mas\,yr$^{-1}$, consistent with the proper motion of Upper
Scorpius. 

The contrast with respect to the brighter member of the close binary
pair is $\Delta K =8.15 \pm 0.08$ and $\Delta J = 8.06 \pm 0.05$,
indicating that the companion has a color of $J-K=0.90 \pm
0.10$. Given that the total extinction along this line of sight is
only $A_V\sim$1.2 or $E(J-K)\sim$0.2 (Schlegel et al. 1998), the
companion must be intrinsically cool ($J-K\ga$0.7 or spectral type
$\ga$K5; Kraus \& Hillenbrand 2007). Combined with its nonmotion
(which indicates it is likely to be quite distant), we therefore
conclude that the candidate companion is likely a background K or
early M giant, perhaps located in the Milky Way bulge. 

\subsubsection{Wider Potential Companions}
\label{sectWider}

Multi-epoch observing campaigns can be observationally expensive, so
where possible, it is best to use archival data to rule out possible
companions. This is sometimes impossible since many candidate
companions of interest can only be distinguished from their candidate
primary using high-resolution imaging techniques. However, wider
companions can often be resolved in seeing-limited data,
especially for new surveys that have very good spatial resolution
(i.e., $\la$0.5\arcsec\, for UKIDSS images). 

Of the 25 companions with separations of $\ga$3.5\arcsec, 21 were
detected in both the $H$ and $K$ filters by UKIDSS, so we can use
their $H-K$ colors to determine whether they might be associated. UKIDSS
observations of known low-mass members of Upper Sco by Lodieu et
al. (2007) show that most members fainter than $K\sim$13 (i.e. with
spectral type $\ga$M7) have colors of $H-K>0.5$; this limit is
consistent with the typical $H-K$ colors of $\ga$M7 field dwarfs as compiled
in Kraus \& Hillenbrand (2007). As we show in Table 3, none of the
candidate companions with both $H$ and $K$ magnitudes meet this
criterion, so we identify all of them to be unassociated field stars,
most likely in the distant background behind Upper Sco. 

In addition, 17 of the wider candidate companions have counterparts
visible in the USNO-B1.0 digitization of the Palomar Observatory Sky
Survey \citep{Monet03}. Most are not present in the USNO-B1.0
source catalog since its source identification algorithm was extremely
conservative in identifying faint neighbors to bright stars. However,
these sources can be manually identified by visual inspection. In
Table 3, we list the bluest plate (B or R) at which each of the
companions was visible. As we noted above, any true companions should
have spectral types of $\ga$M7, so our compilation of field dwarf
colors (Kraus \& Hillenbrand 2007; Kraus \& Hillenbrand, in prep)
suggests that the expected colors for true companions are $B-K>9$ and
$R-K>6$. The detection limits of the POSS survey were $B\sim$21 and
$R\sim$20, so all 17 sources with counterparts in the B or R plates
must be background stars that are bluer than these limits. In all
cases, these identifications agree with the UKIDSS identification. 

Three wider companions in Table~\ref{tblWider} can not be eliminated as Upper Scorpius members
due to insufficient photometry or astrometry. These are the companions
to GSC 06793-00994, GSC 06794-00156 and ScoPMS 015. The closest of
these systems is ScoPMS 015, with a companion just outside of
500\,AU. This small incompleteness defines the outer limit of our
survey until additional follow-up observations can be obtained, though
some azimuthal uncertainty remains at separations of $<$500 AU due to
image boundaries (Section~\ref{sectDectLimits}). 

\begin{deluxetable*}{lrrrcrrr}
 \tabletypesize{\scriptsize}
 \tablewidth{0pt}
 \tablecaption{Faint Candidate Companions to Young Stars in Upper Scorpius}
 \tablehead{\colhead{Known Member} & \colhead{$\rho$} & \colhead{PA} & 
\colhead{$\Delta K$} & \colhead{USNO-B1.0} & 
\colhead{$H_{\rm UKIDSS}$} & \colhead{$K_{\rm UKIDSS}$} & \colhead{$H-K$}
\\
\colhead{} & \colhead{(mas)} & \colhead{(deg)} & \colhead{(mag)} & 
\colhead{(color/epoch)} & \colhead{(mag)} & \colhead{(mag)} & \colhead{(mag)}
 }
 \startdata
RXJ1603.6-2245&10959$\pm$34&206$\pm$0.2&7.52$\pm$0.03&R2&15.339$\pm$0.011&15.074$\pm$0.013&0.265$\pm$0.017\\
RXJ1603.9-2031A&9484$\pm$29&280$\pm$0.2&7.49$\pm$0.04&B2&15.668$\pm$0.012&15.547$\pm$0.018&0.121$\pm$0.022\\
RXJ1606.2-2036&6820$\pm$22&62.6$\pm$0.2&7.07$\pm$0.07&..&16.449$\pm$0.022&16.306$\pm$0.034&0.143$\pm$0.040\\
RXJ1607.0-2036&11939$\pm$36&47.7$\pm$0.2&6.50$\pm$0.03&B2&15.573$\pm$0.010&15.433$\pm$0.015&0.140$\pm$0.018\\
USco-160517.9-202420&7638$\pm$4&153.998$\pm$0.013&6.23$\pm$0.02&B2&15.299$\pm$0.009&15.173$\pm$0.013&0.126$\pm$0.016\\
USco-160801.4-202741&7175$\pm$24&323.8$\pm$0.2&6.88$\pm$0.11&..&16.344$\pm$0.029&16.145$\pm$0.039&0.199$\pm$0.049\\
USco-160801.4-202741&10666$\pm$32&69.1$\pm$0.2&6.67$\pm$0.04&B2&16.053$\pm$0.022&15.959$\pm$0.033&0.094$\pm$0.040\\
USco-160900.7-190852&6850$\pm$27&274.2$\pm$0.2&7.06$\pm$0.05&B2&15.571$\pm$0.012&15.406$\pm$0.017&0.165$\pm$0.021\\
GSC 06205-00954&15398$\pm$47&183.7$\pm$0.2&5.03$\pm$0.02&B2&13.887$\pm$0.003&13.723$\pm$0.004&0.164$\pm$0.005\\
GSC 06209-01501&8994$\pm$27&216.6$\pm$0.2&8.17$\pm$0.04&B2&16.513$\pm$0.033&16.51$\pm$0.055&0.003$\pm$0.064\\
GSC 06213-01358\tablenotemark{a}&4261$\pm$14&219.5$\pm$0.2&8.63$\pm$0.04&..&..&..&..\\
GSC 06793-00797&11086$\pm$33&116.4$\pm$0.2&6.46$\pm$0.02&B2&14.426$\pm$0.005&14.339$\pm$0.008&0.087$\pm$0.009\\
GSC 06793-00797&12330$\pm$38&223.3$\pm$0.2&4.46$\pm$0.03&B2&12.265$\pm$0.001&12.121$\pm$0.001&0.144$\pm$0.001\\
GSC 06793-00994&5462$\pm$17&357.4$\pm$0.2&7.79$\pm$0.04&..&..&15.400$\pm$0.018&..\\
GSC 06794-00480&11820$\pm$36&313.0$\pm$0.2&6.41$\pm$0.03&B2&15.072$\pm$0.008&14.726$\pm$0.010&0.346$\pm$0.013\\
GSC 06214-00210&12941$\pm$39&30.4$\pm$0.2&5.88$\pm$0.02&B2&14.507$\pm$0.006&14.415$\pm$0.008&0.092$\pm$0.010\\
GSC 06794-00537&15942$\pm$49&81.5$\pm$0.2&8.29$\pm$0.06&..&16.310$\pm$0.023&15.951$\pm$0.030&0.359$\pm$0.038\\
GSC 06794-00537&5214$\pm$16&73.6$\pm$0.2&7.99$\pm$0.04&..&15.548$\pm$0.012&15.433$\pm$0.019&0.115$\pm$0.022\\
GSC 06794-00156&5973$\pm$18&338.7$\pm$0.2&9.43$\pm$0.04&..&..&15.029$\pm$0.013&..\\
ScoPMS 015&16954$\pm$51&180.8$\pm$0.2&8.02$\pm$0.07&B2&16.664$\pm$0.033&16.491$\pm$0.044&0.173$\pm$0.055\\
ScoPMS 015&3538$\pm$11&94.9$\pm$0.2&7.19$\pm$0.02&..&..&..&..\\
ScoPMS 015&7485$\pm$23&25.7$\pm$0.2&7.20$\pm$0.03&B2&16.085$\pm$0.020&16.054$\pm$0.030&0.031$\pm$0.036\\
ScoPMS 015&12482$\pm$38&47.0$\pm$0.2&7.06$\pm$0.04&B2&16.773$\pm$0.036&16.618$\pm$0.049&0.155$\pm$0.061\\
ScoPMS 045&7063$\pm$21&190.3$\pm$0.2&8.44$\pm$0.04&B2&16.408$\pm$0.029&16.134$\pm$0.029&0.274$\pm$0.041\\
ScoPMS 045&11286$\pm$34&173.5$\pm$0.2&7.76$\pm$0.03&B2&16.216$\pm$0.024&15.970$\pm$0.025&0.246$\pm$0.035\\
 \enddata
\tablenotetext{a}{See section \ref{sectPFifty} for multi-epoch
  astrometry of this object.}
\label{tblWider}
\end{deluxetable*}

\subsection{Detection Limits}
\label{sectDectLimits}

In order to infer the properties of the distribution of wide ULMCs, it
is essential not only to establish the existance of a small number of
physical companions, but also to determine the magnitude limit as a
function of separation for possible companions not confidently
identified in the observations. This has not been general practice in
previously reported wide ULMCs, in particular not for several of the 
$\la$40\,M$_J$ companions in our separation and primary
mass range: GQ~Lup~b, CT~Cha~b and \pfifty~b. In this section, we will
describe the detection limits for all stars in our sample.

As we described in Section \ref{sectDisco}, we measured our detection
limits in a method that accounts for both spurious detections from
speckle noise (at small separations) and the sky background limit (at
large separations).
These limits could in principle be used as input for Monte Carlo or
Bayesian techniques to study the underlying population, though as we
discuss in the Section \ref{sectFreq}, using the results from our
survey alone could yield a biased measurement.

\begin{figure}
\epsscale{0.8}
\plotone{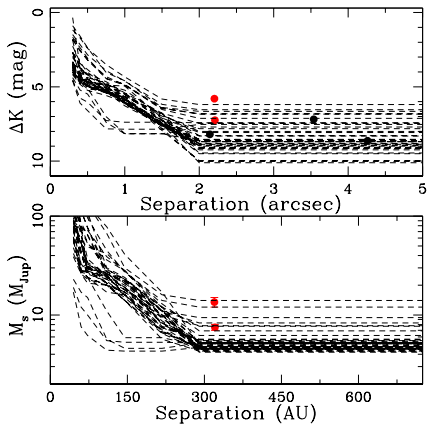}
\caption{Detections and detection limits for our direct imaging observations of
young stars in Upper Sco. Top: Contrast limits ($\Delta K$ in mag) as
a function of angular separation (in arcseconds). Bottom:
Corresponding limits in terms of secondary mass (in $M_{Jup}$) and
physical separation (in AU). The detection limits are shown with b
lack dashed lines, the two confirmed ULMCs are
shown with red points, and all other candidate companions (most of
which are confirmed as background stars) are shown with black
points. We have truncated the plot at a maximum separation of 5\arcsec
since deep seeing-limited imaging from UKIDSS demonstrates that all
wider companions are unassociated field stars (Section 3.2.2).} 
\label{figLimits}
\end{figure}

Table~\ref{tabLimits} lists and Figure~\ref{figLimits} shows 
the companion detection limits for each
star in the sample. The last column of Table~\ref{tabLimits} also
gives the maximum separation in arcseconds where we surveyed all position
angles for companions. The few small ($\la$3\arcsec) values of $\rho_{\rm 100\%}$ in
this column are due to quick image sets taken in the camera sub-array
mode we used for aperture-masking interferometry.
The limits at separations smaller than 1\arcsec\,
are set by the separation-dependent speckle noise and extended PSF
halo of the primary star, while the wider limits at separation greater
than 1.5\arcsec are constant and result from the sky background (for
broadband $K$ observations) or read noise (for narrowband $Br\gamma$
observations).
For each star, we list the primary $K$ magnitude (from 2MASS, with the
flux from any close binary companions subtracted), the detection limit
in $K_{sec}$ at a range of angular separation, and the corresponding
detection limit in $M_{Jup}$ at the corresponding projected orbital
distances. The  mass detection limits were derived from the K
magnitudes of the 5 Myr
DUSTY isochrone \citep{Chabrier00}.

\begin{deluxetable*}{lrrrrrrrcccccccc}
\tabletypesize{\scriptsize}
\tablewidth{0pt}
\tablecaption{Detection limits for additional companions}
\tablehead{\colhead{Name} &
\multicolumn{7}{c}{$K_{lim}$ (mag) at $\rho=$ (mas)} &
\multicolumn{7}{c}{$M_{lim}$ ($M_{Jup}$) at $\rho=$ (AU)} &
\colhead{$\rho_{\rm 100\%}$}
\\
\colhead{} &
\colhead{300} & \colhead{400} & \colhead{500} & \colhead{750} &
\colhead{1000} & \colhead{1500} & \colhead{$\ge$2000} &
\colhead{45} & \colhead{60} & \colhead{75} & \colhead{110} &
\colhead{150} & \colhead{225} &  \colhead{$\ge$300} &\colhead{(\arcsec)}
}
\startdata
GSC 06205-00954&12.9&13.4&13.4&14.1&15.0&16.4&17.8&42 &28 &28 &21 &14&6.9&4.5&10.4\\
GSC 06208-00834&12.6&13.5&13.6&14.0&14.8&16.4&17.5&57 &27 &26 &22 &16&6.8&4.7&10.7\\
GSC 06209-01501&12.3&13.3&13.1&13.7&14.8&16.1&17.7&74 &30 &33 &25 &16&7.6&4.5&10.4\\
GSC 06213-00194&12.1&12.7&13.3&13.6&14.0&15.7&17.1&78 &52 &30 &26 &22&11&5.4&10.2\\
GSC 06213-01358&12.9&13.6&13.8&14.2&15.2&16.5&17.9&39 &26 &23 &20 &14&6.5&4.4&10.4\\
GSC 06214-00210&12.5&13.3&13.8&14.6&14.8&16.0&17.4&62 &29 &24 &17 &16&8.4&4.8&9.4\\
GSC 06214-02384&12.3&12.9&13.5&13.9&14.4&15.8&17.5&76 &41 &27 &23 &18&9.9&4.8&10.2\\
GSC 06764-01305&12.4&13.1&13.6&14.1&14.8&16.2&16.9&71 &33 &27 &21 &16&7.4&5.7&11.1\\
GSC 06793-00797&11.7&12.9&13.2&13.8&14.4&16.0&17.3&97 &42 &31 &24 &19&8.3&5.0&10.2\\
GSC 06793-00994&12.1&12.8&13.4&13.5&14.1&15.7&17.2&78 &43 &29 &28 &21&10&5.2&10.2\\
GSC 06794-00156& 9.8&10.7&11.4&12.1&12.9&14.6&17.2&413&212&130&78 &41&17&5.1&10.2\\
GSC 06794-00480&12.0&12.7&13.1&13.8&13.9&16.0&17.0&81 &53 &33 &24 &23&8&5.5&10.2\\
GSC 06794-00537&11.9&12.7&13.1&13.6&14.0&15.7&17.7&89 &54 &34 &26 &22&10&4.6&10.2\\
RXJ1550.0-2312 &13.7&14.4&15.1&16.8&17.4&17.2&17.1&25 &19 &14 &5.8&4.9&5.1&5.4&2.1\\
RXJ1550.9-2534 &11.4&12.8&13.8&15.6&16.8&17.1&17.3&129&44 &24 &11 &5.9&5.4&4.9&1.3\\
RXJ1551.1-2402 &14.3&15.5&15.8&17.0&17.1&17.1&17.2&19 &12 &10 &5.5&5.3&5.3&5.2&1.5\\
RXJ1557.8-2305 &13.7&15.0&15.5&16.9&17.3&17.2&17.1&25 &15 &12 &5.7&4.9&5.1&5.4&3.7\\
RXJ1558.1-2405 &11.4&12.5&13.2&14.1&14.6&16.1&17.0&126&64 &32 &21 &17&7.7&5.6&8.9\\
RXJ1558.2-2328 &11.4&12.1&12.7&13.3&13.9&15.4&17.1&128&78 &54 &29 &23&12&5.2&10.2\\
RXJ1600.7-2127 &12.5&13.6&14.1&14.2&15.0&16.3&17.6&61 &26 &21 &20 &15&7.1&4.6&10.6\\
RXJ1601.1-2113 &12.7&13.4&13.5&13.7&14.8&16.5&17.4&56 &28 &27 &25 &16&6.6&4.8&9.6\\
RXJ1601.9-2008 &11.7&12.2&12.3&12.9&13.5&15.6&17.6&101&76 &72 &39 &28&11&4.6&11\\
RXJ1602.0-2221 &13.1&14.4&15.1&16.8&16.9&17.0&17.1&35 &19 &14 &5.8&5.7&5.5&5.3&3.7\\
RXJ1602.8-2401A&11.1&11.8&12.4&14.1&15.0&16.0&17.0&156&95 &71 &20 &15&8.4&5.6&3.6\\
RXJ1602.8-2401B&10.1&10.6&11.3&12.4&13.3&14.7&15.1&332&226&131&67 &30&16&14.0&10.2\\
RXJ1603.6-2245 &11.9&12.7&13.4&13.8&14.4&15.8&17.4&89 &54 &28 &24 &19&9.3&4.8&10.2\\
RXJ1603.9-2031A&11.9&12.7&13.4&13.6&14.5&16.0&17.6&88 &54 &29 &27 &18&8.4&4.6&10.1\\
RXJ1604.3-2130 &11.9&12.3&13.0&13.7&14.3&15.9&17.1&83 &73 &37 &25 &19&8.7&5.2&8.9\\
RXJ1606.2-2036 &10.4&11.1&11.7&12.8&13.5&15.1&15.4&271&164&96 &45 &27&14&12.0&11.1\\
RXJ1607.0-2036 & 9.1&12.4&13.3&13.9&14.5&15.6&17.9&648&69&30&23&18&11&4.3&8.7\\
ScoPMS015&11.3 &13.4&13.8&14.1&15.1&16.9&17.9&140&28&24&20&14&5.7&4.3&10.4\\
ScoPMS017&13.9 &14.7&15.0&16.9&17.6&17.4&17.2&23&17&15&5.7&4.6&4.9&5.2&1.5\\
ScoPMS019&11.3 &12.1&12.5&13.5&13.9&15.5&16.6&133&79&65&27&23&11&6.2&8.8\\
ScoPMS022&14.3 &15.5&15.5&17.4&17.4&17.3&17.3&19&12&12&4.9&4.9&4.9&5.0&1.4\\
ScoPMS027&11.8 &12.6&12.7&12.9&13.5&15.3&16.4&89&57&49&42&28&13&6.9&3.0\\
ScoPMS028&13.1 &15.0&15.9&17.1&17.2&17.0&16.9&34&15&9.3&5.3&5.1&5.4&5.7&3.7\\
ScoPMS044&11.8 &12.3&12.7&13.0&13.4&15.5&17.7&91&72&53&36&28&12&4.5&10.3\\
ScoPMS045&12.8 &13.4&13.8&14.2&14.6&15.8&18.0&48&28&24&20&17&9.6&4.2&10.7\\
USco-160341.8-200557&14.0&15.4&15.7&17.3&17.4&17.4&17.3&22 &12 &10 &5 &4.8&4.9&5.0&2.2\\
USco-160643.8-190805&10.1&10.8&11.6&13.2&13.8&15.6&15.8&336&203&110&30&24&11&9.4&8.5\\
USco-160707.7-192715&14.6&16.1&16.6&17.9&17.9&17.7&17.6&17 &7.6&6.2&4.4&4.3&4.5&4.7&1.5\\
USco-160801.4-202741&11.1&11.9&12.5&13.7&14.1&15.8&16.1&158&86 &64 &25&20&10&7.8&8.5\\
USco-160823.2-193001&14.4&15.5&15.7&17.3&17.4&17.3&17.3&18 &11 &11 &5&4.9&5.0&5.0&3.7\\
USco-160825.1-201224&14.3&15.8&16.3&17.3&17.5&17.5&17.5&19 &10 &7  &4.9&4.8&4.8&4.8&1.5\\
USco-160900.7-190852&10.8&11.5&12.2&13.6&14.1&15.6&16.0&196&115&76 &26&21&11&8.3&6.0\\
USco-160916.8-183522&14.3&15.8&16.0&17.4&17.7&17.6&17.4&19 &9.6&8.1&4.9&4.5&4.7&4.9&2.1\\
USco-160954.4-190654&14.5&15.5&16.0&17.4&17.4&17.5&17.5&18 &12 &8.4&4.9&4.9&4.8&4.7&2.2\\
USco-161031.9-191305&10.8&11.5&12.2&13.3&13.9&15.5&16.1&200&114&77 &29&23&12&7.7&8.5\\
USco-161347.5-183459&13.4&13.5&13.5&13.4&13.3&13.2&13.1&28 &27 &27 &29&29&31&34&2.1\\
\enddata
\tablecomments{Detection limits are 5-$\sigma$ for all columns except
  for the $\ge$2000\,mas and $\ge$300\,AU columns, where they are
  10-$\sigma$. The 5\,Myr isochrones of the DUSTY models were used to
  compute K magnitudes. $\rho_{100\%}$ refers to the maximum separation where
  our survey is 100\% complete for additional companions.}
\label{tabLimits}
\end{deluxetable*}

\section{The Frequency of Wide Ultra-Low Mass Companions}
\label{sectFreq}

The most basic
step in assessing the ability of current planet formation theories
(see Section~1) to explain the widely separated and relatively 
massive candidate planetary companions that are observed is to measure
their frequency.
Most of the 49 stars in our sample
have relatively uniform detection limits ($\sim$4--6 $M_{Jup}$ at
$\ga$300\,AU), so a naive estimate of the frequency is approximately
2/49=4.1$^{+4.9}_{-1.3}$\%. Here we have quoted the most likely
frequency, and the Bayesian 68\% confidence interval on the frequency
with a prior distribution where all frequencies are equally likely.
The detection limits are not completely
uniform, so the exact frequency will depend on more sophisticated
analysis using Monte Carlo (K08) or Bayesian (\citealt{Allen07}; 
Kraus and Hillenbrand, in prep.) techniques, and ultimately
should depend on the separation and mass distributions of ULMCs. 

There is also a more fundamental issue that must be
considered: this survey is not the first to be sensitive to the
presence of ULMCs, and a full treatment
should consider all surveys that have properly reported null
detections and detection limits
\citep[e.g.][]{Sartoretti98,Massarotti05,Tanner07,Lafreniere08b,Metchev09,Chauvin10}.
\citet{Chauvin10} reported 1 detection (AB~Pic) from 30-40
young solar-type targets surveyed (depending on the definitions of
``young'' and ``solar-type'').
Indeed, the full sample of past null
detections is larger than our observed sample, suggesting that the
true frequency could be lower by up to a factor of $\ga$2, and that our survey had
good fortune to discover two new companions. Alternatively, including the
companions to CT~Cha and possibly GQ~Lup (likely more massive
than 20\,$M_J$) may increase the true frequency of wide ULMCs, if only
the details of the survey samples in which these companions were
discovered were known.
A large census of the literature is beyond the scope of a discovery
paper, and the results of this analysis will be reported in a
companion paper (Kraus et al., in prep.).

\section{Discussion: Implications for Formation Mechanism}
\label{sectDiscuss}

The population of directly imaged wide ($\ga$40\,AU) ULMCs poses a
significant challenge to planet formation models. Existing models of
solar-system scale planet formation (i.e., via core accretion or gravitational instability
in a Class II disk; \citealt{Pollack96}) have not been succesful at
forming planets at the orbital radii they are
observed here ($\ga$200 AU). It is also difficult to form wide ULMCs
like a binary (i.e. via fragmentation of the freefalling protostellar core or
fragmentation in the massive protostellar disk) without subsequently
accreting sufficient mass to become a stellar or brown dwarf
companion. However, one of these mechanisms must occur. Complicating
this picture is the possibility, and indeed likelihood, of multiple
planet scattering in some formation scenarios. Therefore, we cannot answer the
question of formation mechanism without examining the population of
ULMCs in separation and mass space alongside more well studied
classes of companions.

The core-accretion mechanism for planetary formation only operates
close to the host star ($\sim$5\,AU), where protoplanetary disk densities are high
enough to enable dust and ice to coagulate into
protoplanets. Therefore, a wide ULMC could only originate from
core-accretion if it came from a scattering event. In any
scattering event, low-mass companions are preferentially scattered
outwards, so a core-accretion and scattering origin for wide ULMCs
would require an even greater population of close, higher-mass
objects. As brown dwarfs are well known to be rare less than
$\sim$5\,AU from their host stars \citep[the ``brown-dwarf desert'',
][]{Marcy00}, this scenario is very unlikely.

Another plausible formation scenario for wide ULMCs appears to be
formation like stars from the fragmentation of protostellar clouds and
massive circumstellar disks (e.g. \citealt{Kratter10}). In existing
models, it is difficult to form ULMCs because fragments subsequently
accrete mass, and the only fragments that remain near the opacity
limit for fragmentation are those that are ejected \citep{Bate02}.
In an extreme case, 
the fragmentation of each core might be expected to be independent and companions
would be expected to follow an initial-mass function (i.e. random
pairing). This is now well known to be incorrect, with an
approximately linearly flat distribution of companion masses for solar
type primaries and $\sim$10$^2$\,AU separations
(\citealt{Kraus08}; \citealt{Raghavan10}; Kraus et al 2010 in prep.).
As young solar-type stars have stellar
companion fractions in the range 12--22\%
(\citealt{Brandner96}; K08; Kraus et al, 2010, submitted) per
decade of separation, this would mean that 0.2-0.3\% of solar-type stars
should have $\sim$50-500\,AU, $q=$0.006 to 0.02 companions
(i.e. 6--20\,$M_J$ for solar-type stars. We see a
clear surplus to this model, strongly suggesting that wide ULMCs follow
a different formation path to stars.
 
If wide ULMCs were to form via fragmentation of a 
circumstellar disk when the primary is at the Class~II stage (disk
masses 0.001-0.1\,M$_{\odot}$, e.g. \citealt{Boss01}) it would be much
easier to form wide companions 
than with core accretion. Although most observed disks have masses of
$\sim$5\,M$_J$, a few disks around solar-type stars
(e.g. DL~Tau) have large enough linear dimensions and mass (up to
$\sim$1000\,AU and 0.1\,$M_\sun$,
\citealt{Andrews05,Andrews07}) to fragment into observed 
ULMCs. Relatively little work has been done examining in detail how
these large disks might fragment, with the bulk of disk-fragmentation
literature discussing the possible formation of solar-system scale
planets via fragmentation. Where models of large disks have been
computed (e.g. \citealt{Meru10}), it is clear that it is easier for
them to become Toomre-unstable and fragment than $\sim$20\,AU
disks. The remaining questions to be answered about this fragmentation
include how many fragments are expected, and if the fragments
can accrete enough of the disk mass to become objects like \pseventy~b and \pfifty~b.

The question of where each formation mechanism operates is certainly
not answered yet, but could be in the next few years. Radial velocity
techniques will provide real constraints on the $\sim$5-10\,AU giant
planet frequency as their time baselines increase, while direct
imaging surveys will finish probing the wide ULMC frequency around
young stars. Finally, high angular resolution techniques
such as aperture-masking and high-efficiency coronography will fill
much of the yet-unprobed $\sim$5-30\,AU regime of orbital
separations which likely forms the boundary between the dominance of
core-accretion and fragmentation processes.

\section*{Acknowledgments}
M.I. would like to acknowledge support from the Australian Research
Council through an Australian Postdoctoral Fellowship. We would like
to express thanks to undergraduate students Alison Hammond and Matthew
Hill from the University of Sydney who made a first-pass
astrometric analysis of the data. 
ALK was
suported by a SIM Science Study and by NASA through Hubble Fellowship
grant 51257.01 awarded by the Space Telescope Science Institute, which
is operated by the Association of Universities for Research in
Astronomy, Inc., for NASA, under contract NAS 5-26555.
Some of these observations were
obtained at the Hale Telescope at Palomar
Observatory, as part of a collaborative agreement between
the California Institute of Technology, JPL and Cornell
University. Some of the data presented herein 
were obtained at the W.M. Keck Observatory, which is operated as a
scientific partnership among the California Institute of Technology,
the University of California and the National Aeronautics and Space
Administration. The Observatory was made possible by the generous
financial support of the W.M. Keck Foundation. The authors wish to
recognize and acknowledge the very significant cultural role and
reverence that the summit of Mauna Kea has always had within the
indigenous Hawaiian community.  We are most fortunate to have the
opportunity to conduct observations from this mountain.  


\begin{thebibliography}{65}
\expandafter\ifx\csname natexlab\endcsname\relax\def\natexlab#1{#1}\fi

\bibitem[{{Allen}(2007)}]{Allen07}
{Allen}, P.~R. 2007, \apj, 668, 492

\bibitem[{{Allers} {et~al.}(2010){Allers}, {Liu}, {Dupuy}, \&
  {Cushing}}]{Allers10}
{Allers}, K.~N., {Liu}, M.~C., {Dupuy}, T.~J., \& {Cushing}, M.~C. 2010, \apj,
  715, 561

\bibitem[{{Anderson} {et~al.}(2010){Anderson}, {Collier Cameron}, {Hellier},
  {Lendl}, {Maxted}, {Pollacco}, {Queloz}, {Smalley}, {Smith}, {Todd},
  {Triaud}, {West}, {Barros}, {Enoch}, {Gillon}, {Lister}, {Pepe},
  {S{\'e}gransan}, {Street}, \& {Udry}}]{Anderson10}
{Anderson}, D.~R., {et~al.} 2010, submitted to ApJL, ArXiv:1010.3006 

\bibitem[{{Andrews} \& {Williams}(2005)}]{Andrews05}
{Andrews}, S.~M., \& {Williams}, J.~P. 2005, \apj, 631, 1134

\bibitem[{{Andrews} \& {Williams}(2007)}]{Andrews07}
---. 2007, \apj, 659, 705

\bibitem[{{Baraffe} {et~al.}(1998){Baraffe}, {Chabrier}, {Allard}, \&
  {Hauschildt}}]{Baraffe98}
{Baraffe}, I., {Chabrier}, G., {Allard}, F., \& {Hauschildt}, P.~H. 1998, \aap,
  337, 403

\bibitem[{{Baraffe} {et~al.}(2003){Baraffe}, {Chabrier}, {Barman}, {Allard}, \&
  {Hauschildt}}]{Baraffe03}
{Baraffe}, I., {Chabrier}, G., {Barman}, T.~S., {Allard}, F., \& {Hauschildt},
  P.~H. 2003, \aap, 402, 701

\bibitem[{{Bate}(2005)}]{Bate05}
{Bate}, M.~R. 2005, \mnras, 363, 363

\bibitem[{{Bate} {et~al.}(2002){Bate}, {Bonnell}, \& {Bromm}}]{Bate02}
{Bate}, M.~R., {Bonnell}, I.~A., \& {Bromm}, V. 2002, \mnras, 332, L65

\bibitem[{{Bertin} \& {Arnouts}(1996)}]{Bertin1996}
{Bertin}, E., \& {Arnouts}, S. 1996, \aaps, 117, 393

\bibitem[{{Boss}(2001)}]{Boss01}
{Boss}, A.~P. 2001, \apj, 563, 367

\bibitem[{{Bouchy} {et~al.}(2010){Bouchy}, {Deleuil}, {Guillot}, {Aigrain},
  {Carone}, \& {Cochran}}]{Bouchy10}
{Bouchy}, F., {Deleuil}, M., {Guillot}, T., {Aigrain}, S., {Carone}, L., \&
  {Cochran}, W.~D. 2010, accepted for A\&A, ArXiv:1010.0179

\bibitem[{{Brandner} {et~al.}(1996){Brandner}, {Alcala}, {Kunkel}, {Moneti}, \&
  {Zinnecker}}]{Brandner96}
{Brandner}, W., {Alcala}, J.~M., {Kunkel}, M., {Moneti}, A., \& {Zinnecker}, H.
  1996, \aap, 307, 121

\bibitem[{{Cameron}(2008)}]{Cameron08}
{Cameron}, B.~P. 2008, PhD thesis, California Institute of Technology, United
  States -- California

\bibitem[{{Cameron} {et~al.}(2009){Cameron}, {Britton}, \&
  {Kulkarni}}]{Cameron09}
{Cameron}, P.~B., {Britton}, M.~C., \& {Kulkarni}, S.~R. 2009, \aj, 137, 83

\bibitem[{{Chabrier} {et~al.}(2000){Chabrier}, {Baraffe}, {Allard}, \&
  {Hauschildt}}]{Chabrier00}
{Chabrier}, G., {Baraffe}, I., {Allard}, F., \& {Hauschildt}, P. 2000, \apj,
  542, 464

\bibitem[{{Chauvin} {et~al.}(2004){Chauvin}, {Lagrange}, {Dumas}, {Zuckerman},
  {Mouillet}, {Song}, {Beuzit}, \& {Lowrance}}]{Chauvin04}
{Chauvin}, G., {Lagrange}, A., {Dumas}, C., {Zuckerman}, B., {Mouillet}, D.,
  {Song}, I., {Beuzit}, J., \& {Lowrance}, P. 2004, \aap, 425, L29

\bibitem[{{Chauvin} {et~al.}(2010){Chauvin}, {Lagrange}, {Bonavita},
  {Zuckerman}, {Dumas}, {Bessell}, {Beuzit}, {Bonnefoy}, {Desidera}, {Farihi},
  {Lowrance}, {Mouillet}, \& {Song}}]{Chauvin10}
{Chauvin}, G., {et~al.} 2010, \aap, 509, A52+

\bibitem[{{Currie} {et~al.}(2009){Currie}, {Lada}, {Plavchan}, {Robitaille},
  {Irwin}, \& {Kenyon}}]{Currie09}
{Currie}, T., {Lada}, C.~J., {Plavchan}, P., {Robitaille}, T.~P., {Irwin}, J.,
  \& {Kenyon}, S.~J. 2009, \apj, 698, 1

\bibitem[{{de Zeeuw} {et~al.}(1999){de Zeeuw}, {Hoogerwerf}, {de Bruijne},
  {Brown}, \& {Blaauw}}]{deZeeuw99}
{de Zeeuw}, P.~T., {Hoogerwerf}, R., {de Bruijne}, J.~H.~J., {Brown}, A.~G.~A.,
  \& {Blaauw}, A. 1999, \aj, 117, 354

\bibitem[{{Deleuil} {et~al.}(2008){Deleuil}, {Deeg}, {Alonso}, {Bouchy},
  {Rouan}, {Auvergne}, {Baglin}, {Aigrain}, {Almenara}, {Barbieri}, {Barge},
  {Bruntt}, {Bord{\'e}}, {Collier Cameron}, {Csizmadia}, {de La Reza},
  {Dvorak}, {Erikson}, {Fridlund}, {Gandolfi}, {Gillon}, {Guenther}, {Guillot},
  {Hatzes}, {H{\'e}brard}, {Jorda}, {Lammer}, {L{\'e}ger}, {Llebaria},
  {Loeillet}, {Mayor}, {Mazeh}, {Moutou}, {Ollivier}, {P{\"a}tzold}, {Pont},
  {Queloz}, {Rauer}, {Schneider}, {Shporer}, {Wuchterl}, \&
  {Zucker}}]{Deleuil08}
{Deleuil}, M., {et~al.} 2008, \aap, 491, 889

\bibitem[{{Fluks} {et~al.}(1994){Fluks}, {Plez}, {The}, {de Winter},
  {Westerlund}, \& {Steenman}}]{Fluks94}
{Fluks}, M.~A., {Plez}, B., {The}, P.~S., {de Winter}, D., {Westerlund}, B.~E.,
  \& {Steenman}, H.~C. 1994, \aaps, 105, 311

\bibitem[{{Ghez} {et~al.}(1993){Ghez}, {Neugebauer}, \& {Matthews}}]{Ghez93}
{Ghez}, A.~M., {Neugebauer}, G., \& {Matthews}, K. 1993, \aj, 106, 2005

\bibitem[{{Ghez} {et~al.}(2008){Ghez}, {Salim}, {Weinberg}, {Lu}, {Do}, {Dunn},
  {Matthews}, {Morris}, {Yelda}, {Becklin}, {Kremenek}, {Milosavljevic}, \&
  {Naiman}}]{Ghez08}
{Ghez}, A.~M., {et~al.} 2008, \apj, 689, 1044

\bibitem[{{Golimowski} {et~al.}(2004){Golimowski}, {Leggett}, {Marley}, {Fan},
  {Geballe}, {Knapp}, {Vrba}, {Henden}, {Luginbuhl}, {Guetter}, {Munn},
  {Canzian}, {Zheng}, {Tsvetanov}, {Chiu}, {Glazebrook}, {Hoversten},
  {Schneider}, \& {Brinkmann}}]{Golimowski04}
{Golimowski}, D.~A., {et~al.} 2004, \aj, 127, 3516

\bibitem[{{Grether} \& {Lineweaver}(2006)}]{Grether06}
{Grether}, D., \& {Lineweaver}, C.~H. 2006, \apj, 640, 1051

\bibitem[{{Haisch} {et~al.}(2001){Haisch}, {Lada}, \& {Lada}}]{Haisch01}
{Haisch}, Jr., K.~E., {Lada}, E.~A., \& {Lada}, C.~J. 2001, \apjl, 553, L153

\bibitem[{{Hern{\'a}ndez} {et~al.}(2007){Hern{\'a}ndez}, {Hartmann}, {Megeath},
  {Gutermuth}, {Muzerolle}, {Calvet}, {Vivas}, {Brice{\~n}o}, {Allen},
  {Stauffer}, {Young}, \& {Fazio}}]{Hernandez07}
{Hern{\'a}ndez}, J., {et~al.} 2007, \apj, 662, 1067

\bibitem[{{Hillenbrand} \& {White}(2004)}]{Hillenbrand04}
{Hillenbrand}, L.~A., \& {White}, R.~J. 2004, \apj, 604, 741

\bibitem[{{Hoyle}(1953)}]{Hoyle53}
{Hoyle}, F. 1953, \apj, 118, 513

\bibitem[{{Kalas} {et~al.}(2008){Kalas}, {Graham}, {Chiang}, {Fitzgerald},
  {Clampin}, {Kite}, {Stapelfeldt}, {Marois}, \& {Krist}}]{Kalas08}
{Kalas}, P., {et~al.} 2008, Science, 322, 1345

\bibitem[{{Kratter} {et~al.}(2010){Kratter}, {Murray-Clay}, \&
  {Youdin}}]{Kratter10}
{Kratter}, K.~M., {Murray-Clay}, R.~A., \& {Youdin}, A.~N. 2010, \apj, 710,
  1375

\bibitem[{{Kraus}(2009)}]{Krausthesis09}
{Kraus}, A.~L. 2009, PhD thesis, California Institute of Technology, United
  States -- California

\bibitem[{{Kraus} \& {Hillenbrand}(2008)}]{Kraus08b}
{Kraus}, A.~L., \& {Hillenbrand}, L.~A. 2008, \apjl, 686, L111

\bibitem[{{Kraus} \& {Hillenbrand}(2009)}]{Kraus09}
---. 2009, \apj, 703, 1511

\bibitem[{{Kraus} {et~al.}(2008){Kraus}, {Ireland}, {Martinache}, \&
  {Lloyd}}]{Kraus08}
{Kraus}, A.~L., {Ireland}, M.~J., {Martinache}, F., \& {Lloyd}, J.~P. 2008,
  \apj, 679, 762

\bibitem[{{Kunkel}(1999)}]{Kunkel99}
{Kunkel}, M. 1999, PhD thesis, University of Wurzburg

\bibitem[{{Lafreni{\`e}re} {et~al.}(2008{\natexlab{a}}){Lafreni{\`e}re},
  {Jayawardhana}, {Brandeker}, {Ahmic}, \& {van Kerkwijk}}]{Lafreniere08b}
{Lafreni{\`e}re}, D., {Jayawardhana}, R., {Brandeker}, A., {Ahmic}, M., \& {van
  Kerkwijk}, M.~H. 2008{\natexlab{a}}, \apj, 683, 844

\bibitem[{{Lafreni{\`e}re} {et~al.}(2008{\natexlab{b}}){Lafreni{\`e}re},
  {Jayawardhana}, \& {van Kerkwijk}}]{Lafreniere08}
{Lafreni{\`e}re}, D., {Jayawardhana}, R., \& {van Kerkwijk}, M.~H.
  2008{\natexlab{b}}, \apjl, 689, L153

\bibitem[{{Lafreni{\`e}re} {et~al.}(2010){Lafreni{\`e}re}, {Jayawardhana}, \&
  {van Kerkwijk}}]{Lafreniere10}
---. 2010, \apj, 719, 497

\bibitem[{{Lafreni{\`e}re} {et~al.}(2007){Lafreni{\`e}re}, {Doyon}, {Marois},
  {Nadeau}, {Oppenheimer}, {Roche}, {Rigaut}, {Graham}, {Jayawardhana},
  {Johnstone}, {Kalas}, {Macintosh}, \& {Racine}}]{Lafreniere07}
{Lafreni{\`e}re}, D., {et~al.} 2007, \apj, 670, 1367

\bibitem[{{Lawrence} {et~al.}(2007){Lawrence}, {Warren}, {Almaini}, {Edge},
  {Hambly}, {Jameson}, q{Lucas}, {Casali}, {Adamson}, {Dye}, {Emerson},
  {Foucaud}, {Hewett}, {Hirst}, {Hodgkin}, {Irwin}, {Lodieu}, {McMahon},
  {Simpson}, {Smail}, {Mortlock}, \& {Folger}}]{Lawrence07}
{Lawrence}, A., {et~al.} 2007, \mnras, 379, 1599

\bibitem[{{Leggett}(1992)}]{Leggett92}
{Leggett}, S.~K. 1992, \apjs, 82, 351

\bibitem[{{Leggett} {et~al.}(2002){Leggett}, {Golimowski}, {Fan}, {Geballe},
  {Knapp}, {Brinkmann}, {Csabai}, {Gunn}, {Hawley}, {Henry}, {Hindsley},
  {Ivezi{\'c}}, {Lupton}, {Pier}, {Schneider}, {Smith}, {Strauss}, {Uomoto}, \&
  {York}}]{Leggett02}
{Leggett}, S.~K., {et~al.} 2002, \apj, 564, 452

\bibitem[{{Lodieu} {et~al.}(2007){Lodieu}, {Hambly}, {Jameson}, {Hodgkin},
  {Carraro}, \& {Kendall}}]{Lodieu07}
{Lodieu}, N., {Hambly}, N.~C., {Jameson}, R.~F., {Hodgkin}, S.~T., {Carraro},
  G., \& {Kendall}, T.~R. 2007, \mnras, 374, 372

\bibitem[{{Low} \& {Lynden-Bell}(1976)}]{Low76}
{Low}, C., \& {Lynden-Bell}, D. 1976, \mnras, 176, 367

\bibitem[{{Mamajek}(2005)}]{Mamajek05}
{Mamajek}, E.~E. 2005, \apj, 634, 1385

\bibitem[{{Marcy} \& {Butler}(2000)}]{Marcy00}
{Marcy}, G.~W., \& {Butler}, R.~P. 2000, \pasp, 112, 137

\bibitem[{{Marois} {et~al.}(2008){Marois}, {Macintosh}, {Barman}, {Zuckerman},
  {Song}, {Patience}, {Lafreni{\`e}re}, \& {Doyon}}]{Marois08}
{Marois}, C., {Macintosh}, B., {Barman}, T., {Zuckerman}, B., {Song}, I.,
  {Patience}, J., {Lafreni{\`e}re}, D., \& {Doyon}, R. 2008, Science, 322, 1348

\bibitem[{{Masciadri} {et~al.}(2005){Masciadri}, {Mundt}, {Henning}, {Alvarez},
  \& {Barrado y Navascu{\'e}s}}]{Masciadri05}
{Masciadri}, E., {Mundt}, R., {Henning}, T., {Alvarez}, C., \& {Barrado y
  Navascu{\'e}s}, D. 2005, \apj, 625, 1004

\bibitem[{{Massarotti} {et~al.}(2005){Massarotti}, {Latham}, {Torres}, {Brown},
  \& {Oppenheimer}}]{Massarotti05}
{Massarotti}, A., {Latham}, D.~W., {Torres}, G., {Brown}, R.~A., \&
  {Oppenheimer}, B.~D. 2005, \aj, 129, 2294

\bibitem[{{Meru} \& {Bate}(2010)}]{Meru10}
{Meru}, F., \& {Bate}, M.~R. 2010, \mnras, 858

\bibitem[{{Metchev} \& {Hillenbrand}(2009)}]{Metchev09}
{Metchev}, S.~A., \& {Hillenbrand}, L.~A. 2009, \apjs, 181, 62

\bibitem[{{Monet} {et~al.}(2003){Monet}, {Levine}, {Canzian}, {Ables}, {Bird},
  {Dahn}, {Guetter}, {Harris}, {Henden}, {Leggett}, {Levison}, {Luginbuhl},
  {Martini}, {Monet}, {Munn}, {Pier}, {Rhodes}, {Riepe}, {Sell}, {Stone},
  {Vrba}, {Walker}, {Westerhout}, {Brucato}, {Reid}, {Schoening}, {Hartley},
  {Read}, \& {Tritton}}]{Monet03}
{Monet}, D.~G., {et~al.} 2003, \aj, 125, 984

\bibitem[{{Pollack} {et~al.}(1996){Pollack}, {Hubickyj}, {Bodenheimer},
  {Lissauer}, {Podolak}, \& {Greenzweig}}]{Pollack96}
{Pollack}, J.~B., {Hubickyj}, O., {Bodenheimer}, P., {Lissauer}, J.~J.,
  {Podolak}, M., \& {Greenzweig}, Y. 1996, Icarus, 124, 62

\bibitem[{{Raghavan} {et~al.}(2010){Raghavan}, {McAlister}, {Henry}, {Latham},
  {Marcy}, {Mason}, {Gies}, {White}, \& {ten Brummelaar}}]{Raghavan10}
{Raghavan}, D., {et~al.} 2010, ArXiv: e-prints

\bibitem[{{Sartoretti} {et~al.}(1998){Sartoretti}, {Brown}, {Latham}, \&
  {Torres}}]{Sartoretti98}
{Sartoretti}, P., {Brown}, R.~A., {Latham}, D.~W., \& {Torres}, G. 1998, \aap,
  334, 592

\bibitem[{{Scholz} {et~al.}(2007){Scholz}, {Jayawardhana}, {Wood}, {Meeus},
  {Stelzer}, {Walker}, \& {O'Sullivan}}]{Scholz07}
{Scholz}, A., {Jayawardhana}, R., {Wood}, K., {Meeus}, G., {Stelzer}, B.,
  {Walker}, C., \& {O'Sullivan}, M. 2007, \apj, 660, 1517

\bibitem[{{Simon} {et~al.}(1995){Simon}, {Ghez}, {Leinert}, {Cassar}, {Chen},
  {Howell}, {Jameson}, {Matthews}, {Neugebauer}, \& {Richichi}}]{Simon95}
{Simon}, M., {et~al.} 1995, \apj, 443, 625

\bibitem[{{Stetson}(1987)}]{Stetson87}
{Stetson}, P.~B. 1987, \pasp, 99, 191

\bibitem[{{Tanner} {et~al.}(2007){Tanner}, {Beichman}, {Akeson}, {Ghez},
  {Grankin}, {Herbst}, {Hillenbrand}, {Huerta}, {Konopacky}, {Metchev},
  {Mohanty}, {Prato}, \& {Simon}}]{Tanner07}
{Tanner}, A., {et~al.} 2007, \pasp, 119, 747

\bibitem[{{Teixeira} {et~al.}(2008){Teixeira}, {Ducourant}, {Chauvin},
  {Krone-Martins}, {Song}, \& {Zuckerman}}]{Teixeira08}
{Teixeira}, R., {Ducourant}, C., {Chauvin}, G., {Krone-Martins}, A., {Song},
  I., \& {Zuckerman}, B. 2008, \aap, 489, 825

\bibitem[{{Troy} {et~al.}(2000){Troy}, {Dekany}, {Brack}, {Oppenheimer},
  {Bloemhof}, {Trinh}, {Dekens}, {Shi}, {Hayward}, \& {Brandl}}]{Troy2000}
{Troy}, M., {et~al.} 2000, in Society of Photo-Optical Instrumentation
  Engineers (SPIE) Conference Series, Vol. 4007, ed. {P.~L.~Wizinowich},
  31--40

\bibitem[{{van Leeuwen}(2007)}]{vanLeeuwen07}
{van Leeuwen}, F. 2007, {Hipparcos, the New Reduction of the Raw Data}
  (Hipparcos, the New Reduction of the Raw Data, Astrophysics and Space Science
  Library, Vol.~ 350 20 Springer Dordrecht)

\bibitem[{{Zacharias} {et~al.}(2010){Zacharias}, {Finch}, {Girard}, {Hambly},
  {Wycoff}, {Zacharias}, {Castillo}, {Corbin}, {DiVittorio}, {Dutta}, {Gaume},
  {Gauss}, {Germain}, {Hall}, {Hartkopf}, {Hsu}, {Holdenried}, {Makarov},
  {Martines}, {Mason}, {Monet}, {Rafferty}, {Rhodes}, {Siemers}, {Smith},
  {Tilleman}, {Urban}, {Wieder}, {Winter}, \& {Young}}]{Zacharias10}
{Zacharias}, N., {et~al.} 2010, \aj, 139, 2184

\end{thebibliography}

\end{document}